\documentclass[aps,prb,twocolumn,superscriptaddress,showpacs,nofootinbib]{revtex4-1}
\usepackage{graphicx,lipsum}
\usepackage{amsmath,color}
\usepackage{pgfplots}
\usepackage[export]{adjustbox}
\usepackage[normalem]{ulem}
\usepackage{bm}

\newcommand{\boldw}{{\bf{w}}}
\newcommand{\boldr}{{\bf{r}}}
\newcommand{\bolds}{{\bf{s}}}

\newcommand{\boldv}{{\bf{v}}}
\newcommand\bom{\boldsymbol{\omega}}
\newcommand\bos{\bm{\sigma}}

\def\refeq#1{(\ref{#1})}

\begin{document}

\title{Mesoscale helicity distinguishes Vinen from Kolmogorov turbulence
 in helium~II}
\author{L.~Galantucci}
\author{C. F. Barenghi}
\author{N.G.~Parker}
\author{A.~W.~Baggaley}

\affiliation{Joint Quantum Centre Durham--Newcastle, School of Mathematics,
Statistics and Physics, Newcastle University, Newcastle upon Tyne, NE1 7RU, United Kingdom}


\begin{abstract}
Experiments and numerical simulations show that quantum turbulence
exists in two distinct limiting regimes: 
Kolmogorov turbulence (which shares with
classical turbulence the important property of a cascade of kinetic energy from
large eddies to small eddies) and Vinen turbulence (which is more similar to
a random flow). 
In this work, we define a mesoscale helicity for the superfluid, which, 
tested in numerical experiments,
distinguishes the two turbulent regimes, quantifying the amount
of nonlocal vortex interactions and the orientation
of the vortex lines.
\end{abstract}

\maketitle

\section{Introduction}\label{sec:intro}
  
Vorticity in superfluid liquid helium (helium~II) is not a continuous
and unconstrained
field, as in ordinary (classical) fluids, but consists of thin vortex lines 
\cite{Primer} whose strength
(measured by the velocity circulation $\kappa$) and thickness (\textit{i.e.}
the vortex core radius $a_0$) are held fixed by quantum mechanical constraints.
In this study we focus on turbulent tangles of vortex lines 
generated when liquid helium is stirred  
\cite{Barenghi2014a,SkrbekSreeni2012,Nemirovskii2013}.
Similar tangles of vortex lines can also be created in 
trapped atomic Bose-Einstein
condensates by laser stirring, by shaking the trap \cite{Henn2009} 
or by temperature quenches \cite{lamporesi-etal-2013}. 
The natural question is whether this state of disorder (hereafter
referred to as \textit{quantum turbulence})
is similar to turbulence in ordinary fluids (\textit{classical turbulence}) or not.  

The evidence so far is that quantum turbulence can assume 
two distinct limiting regimes
  \cite{Volovik2003,Walmsley2008,Nemirovskii2019} 
called respectively {\it Kolmogorov} turbulence
and {\it Vinen} turbulence
(see the Section II for a description of their physical properties).
The Kolmogorov regime has been observed in helium experiments
driven by counter-rotating propellers \cite{Tabeling1998},
wind tunnels \cite{Salort2010}, 
towed grids \cite{Smith1993,Zmeev2015}, vibrating grid in $^3$He-B
\cite{Bradley2006,Bradley2011},
or by the injection of vortex rings \cite{Walmsley2008}, 
and has been reproduced in the
numerical simulations of these experiments 
\cite{Baggaley2012structures,Sherwin2012}. 
At large scales this regime
resembles classical turbulence.
The Vinen regime (and its crossover to the Kolmogorov regime)
has been observed in experiments \cite{Walmsley2008} 
and in numerical simulations \cite{Baggaley2012ultra}
of turbulence driven by vortex ring injection;
it has also
been generated in numerical simulations of helium~II
turbulence driven by a small heat flux \cite{Sherwin2012} 
(also known as counterflow superfluid turbulence),
in numerical simulations
of turbulence in trapped atomic Bose-Einstein condensates \cite{Cidrim2017}
(including the thermal quench of a Bose gas \cite{Stagg2016,bland-etal-2018}), 
and in superfluid models of the early universe \cite{Mocz2017}. 
This regime seems to have no classical counterpart.

Our aim is to show that, in helium~II, Kolmogorov turbulence and Vinen 
turbulence differ also in terms of what we call {\it mesoscale helicity}.
Helicity is a property of great importance in classical fluid dynamics,
but in the context of quantum fluids its role, and even its definition,
are still debated. The mesoscale definition of superfluid helicity 
which here we propose 
extends the classical definition of helicity in fluid dynamics 
to the mesoscale description of turbulent vortex lines in helium~II
provided by the Vortex Filament Model (VFM); the VFM is the best
model currently available for turbulent helium~II at the non-zero
temperatures and length scales typical of most turbulence experiments.
Our results shed new light into the different nature of
Kolmogorov and Vinen turbulent flows. 

The plan of the paper is the following. Firstly
we shall review the difference between Vinen and 
Kolmogorov turbulence regimes (Section II) and recall
the mesoscale description of turbulence provided by the VFM 
(Section III). In Section IV we shall introduce the definition of
mesoscale helicity $\mathcal{H}$ for a superfluid. 
To better understand the physical meaning of $\mathcal{H}$, 
in Section V we shall determine $\mathcal{H}$ for simple vortex 
configurations, before measuring it in two distinct turbulent flows
(Section VI). Section VII will discuss how the values of $\mathcal{H}$
distinguish Vinen from Kolmogorov regimes. Section VIII will draw the
conclusions.

\section{Vinen vs Kolmogorov}

Kolmogorov turbulence \cite{Frisch1995}
is characterised by the same energy spectrum 
${\hat E}(k) \sim k^{-5/3}$ observed in classical turbulence 
in its simplest (homogeneous, isotropic, statistically-steady) form. 
The energy spectrum ${\hat E}(k)$ describes the distribution of 
turbulent kinetic energy 
$E=\int_0^{\infty} {\hat E}(k) dk$ over
the length scales $2 \pi/k$ (where $k$ is the wavenumber).  
The $k^{-5/3}$ scaling is interpreted as the 
manifestation of an energy cascade from large to 
small eddies taking place over the inertial range 
$k_D=2 \pi/D \ll k \ll k_{\eta}=2 \pi/\eta$ 
where $D$ is the (large) length scale of the energy injection 
and $\eta$ is the (small) length scale of the viscous dissipation. 
In quantum turbulence, $k_{\eta}$ must be replaced by
$k_{\ell}=2 \pi/\ell$ where $\ell$ is the
average distance between the vortex lines \cite{Barenghi2014b}; 
at length scales shorter
than $\ell$ ($k \gg k_{\ell}$), individual vortex line dynamics 
(such as Kelvin waves and phonon emission) 
becomes significant and leads to a departure from classical 
hydrodynamic turbulence.
Numerical simulations \cite{Nore1997,Araki2002}
reveal that the $k^{-5/3}$ scaling 
observed for $k \ll k_{\ell}$ in superfluids arises from
the partial polarisation of the vortex tangle: vortex lines align
locally, forming energy-containing 
bundles \cite{BaggaleyLaurie2012,Baggaley2012bundles}
which can induce flow at large scales.
Further evidence of classical behaviour arises from the 
temporal decay of the kinetic energy $E(t)$ and the
vortex line density ${\cal{L}}(t)$  
which are observed when the forcing
which sustains the turbulence in a steady-state is removed
\cite{stalp-skrbek-donnelly-1999,varga-babuin-skrbek-2015,gao-etal-2016}:
$E(t) \sim t^{-2}$ and ${\cal{L}}(t) \sim t^{-3/2}$, where 
${\cal{L}}(t)$ is defined as the vortex length per unit volume
at time $t$.

The second limiting
form of turbulence, Vinen turbulence, has different signatures.
Its energy spectrum ${\hat E}(k)$ lacks the $k^{-5/3}$ scaling and may
peak at the intermediate scales around $k_{\ell}$ 
rather than at the large scales $k_D$; at larger wavenumbers
it displays
the characteristic $k^{-1}$ behaviour of an isolated vortex.
The last feature suggests that Vinen turbulence is a random-like flow
with a weak or absent energy cascade \cite{Barenghi2016}
(indeed the velocity correlation functions become negligible for distances
larger than $\ell$ \cite{Stagg2016,Cidrim2017}). 
If the forcing is removed, Vinen turbulence decays more slowly
than Kolmogorov turbulence: 
$E(t) \sim t^{-1}$ and ${\cal{L}}(t) \sim t^{-1}$. 
For the sake of completeness, it is important to note that
in Vinen turbulence observed in counterflow channels, 
one-point turbulent velocity statistics (PDFs) are Gaussian (as in classical
turbulence) or exhibit quantum peculiar power laws, depending on whether the 
measurement region, 
$\Delta$, is $\Delta \gg \ell$ or $\Delta \ll \ell$, respectively 
\cite{LaMantiaSkrbek2014stats,Baggaley2011stats,galantucci-sciacca-2014}.

\section{Vortex Filament Model}

Since our concern is experiments in helium~II at non-zero temperatures, 
we use the Vortex Filament Model (VFM) \cite{Schwarz1988,HanninenBaggaley2014}.
The VFM is based on the separation of length scales
$a_0 \ll \ell \ll D$ typical of experiments, 
where  $a_0 =10^{-10}~\rm m$ is the vortex core 
radius, $\ell \approx 10^{-5}$ to $10^{-3}~\rm m$ 
is the average distance between vortex lines, and
$D \approx 10^{-2}$ to $10^{-1}~\rm m$ is the system size.
In the range of length scales $\ell < \Delta < D$
relevant to turbulence, the VFM describes vortex lines
as three dimensional, oriented, reconnecting space curves,
$\bolds(\xi,t)$, where $\xi$ is arclength, carrying one
quantum $\kappa$ of circulation.  The curves have infinitesimal thickness,
unlike quantum vortices described by more microscopic approaches
such as the Gross Pitaevskii equation (GPE)
\cite{Primer} or N-body quantum mechanics \cite{GalliReattoRossi2014}; 
in other words, the VFM is a {\it mesoscopic} model which neglects
density variations at the scale of the vortex core. 
At each time $t$, the VFM describes the vortex tangle as a collection 
of $N_v$ oriented curves
of length $L_i$ ($i=1,\cdots N_v$), 
where $N_v$, $L_i$ and the total vortex length
$L=\sum_{i=1}^{N_v} L_i$ vary with $t$.
The velocity field $\boldv(\boldr)$ which all vortex lines induce at a point
$\boldr \neq \bolds$ (i.e. a point which is not on a vortex line)
 is given by the Biot-Savart law \cite{Saffman1993}:

\begin{equation}
\boldv(\boldr,t)=
\frac{\kappa}{4\pi}\int_{\mathcal{T}} d\xi
\frac{\bolds'(\xi,t) \times [\boldr-\bolds(\xi,t)]}{\left|\boldr-\bolds(\xi,t)\right|^3},
\label{eq:BS1}
\end{equation}

\noindent
where the line integral extends over the entire vortex configuration 
$\mathcal{T}$ and $\bolds'=\partial \bolds/\partial \xi$ is the unit tangent 
vector at the point $\bolds$.  At non-zero temperature, the motion of 
vortex lines is determined by Schwarz's equation \cite{Schwarz1988}:
\vspace{-3mm}

\begin{align}
\frac{d \bolds}{dt}&=\boldv(\bolds,t) + \alpha \bolds' \times \boldw(\bolds,t)
- \alpha' \bolds' \times 
[ \bolds' \times \boldw(\bolds,t) ],
\label{eq:Schwarz}
\end{align}

\noindent
where $\boldw(\bolds,t)=\boldv_n(\bolds,t) - \boldv(\bolds,t)$, 
$\boldv_n(\bolds,t)$ is the velocity of the normal fluid at $\bf s$,
and $\alpha$ and $\alpha'$ are temperature-dependent friction 
coefficients \cite{BarenghiDonnelly1998} accounting for the interaction
between normal fluid and vortex lines.
Without a core structure,
the Biot-Savart integral, Eq.~\refeq{eq:BS1}, would diverge if 
evaluated at a point $\boldr=\bolds$ on a vortex line,
and must be desingularized in a standard way \cite{Schwarz1988} 
which takes into account the vortex core's finite size, $a_0$,
and the minimum Lagrangian discretization 
along the vortex lines, $\Delta \xi$.  The total superfluid velocity $\boldv$ 
of the vortex line at the point
$\bolds(\xi,t)$ can thus be decomposed into near, far and external 
contributions, \textit{i.e.}
$\boldv(\bolds,t)= \boldv^{far}(\bolds,t) + \boldv^{near}(\bolds,t) 
+\boldv^{ext}(\bolds,t)$.
The far contribution is

\begin{equation}
\boldv^{far}(\bolds(\xi,t),t)=
\frac{\kappa}{4\pi}\int_{\mathcal{T}'} d\xi'
\frac{\bolds'(\xi',t) \times [\bolds(\xi,t)-\bolds(\xi',t)]}{\left|\bolds(\xi,t)-\bolds(\xi',t)\right|^3},
\label{eq:BS2}
\end{equation}

\noindent
where now the integral extends to all vortex lines
present in the tangle (the line through the point $\bolds$ as well as all
other vortex lines) but excludes the neighborhood of the point $\bolds$
(this is the meaning of the symbol $\mathcal{T}'$).
The near contribution, accounting for this neighborhood, is 

\begin{equation}
\boldv^{near}(\bolds,t)=\frac{\kappa}{4 \pi} \ln{\left( \frac{R_c}{a_0} 
\right)} \bolds' \times \bolds''.
\label{eq:loc}
\end{equation}

\noindent
Note that $\boldv^{near}(\bolds,t)$ is directed in the binormal 
direction and (neglecting the dependence on the
slow varying logarithmic term) is  
inversely proportional to the local radius of curvature 
$R_c=1/\vert \bolds''\vert$ at the point $\bolds$, 
where $\bolds''=\partial^2\bolds/\partial\xi^2$.
Finally the external
contribution, $\boldv^{ext}(\bolds,t)$, describes any 
irrotational flow which is externally applied, for example by bellows
or a heater.

\section{Mesoscale Helicity}

The classical definition of helicity $\mathcal{H}$ is \cite{Moffatt1969}  

\begin{equation}
\mathcal{H}=\iiint_V \bom (\boldr) \cdot \boldv(\boldr) \; d^3\!\boldr,
\label{eq:H1}
\end{equation}

\noindent
where $\boldv$ is the velocity field,
$\bom=\nabla \times \boldv$ is the vorticity field, $\boldr$ is the position
and $V$ is the volume containing the fluid. 
An ideal fluid evolving without viscosity according to the classical
Euler equation conserves both energy and helicity; helicity conservation
freezes the flow's topology in time.  
In real fluids, viscous dissipation destroys both energy and helicity.
Helicity is important in turbulence:
a large value of $\mathcal{H}$ weakens the nonlinearity of the governing
Navier-Stokes equation, reducing the direct cascade of energy from 
large to small length scales \cite{Kraichnan1973}. 
Spontaneous reflectional symmetry breaking, implying nonzero net helicity,
has been reported in
turbulent flows with initial and boundary conditions which are symmetric
\cite{Levich,Kholmyansky}.
Recent work \cite{Biferale2012} has also shown that the interaction 
of helical modes of the same sign favours three-dimensional
inverse energy transfer and the generation of large
length scales.  In astrophysics, helicity 
quantifies the lack of mirror symmetry of the flow which favours 
the generation of magnetic field by dynamo action 
\cite{BrandenburgSubramanian2005}. 
It is now possible to directly measure the helicity of thin-cored
vortices \cite{Scheeler2017},
boosting the interest in the role kinetic helicity plays in
constraining the hydrodynamics.

The superfluid component of helium~II has zero viscosity and
can thus be considered as the closest physical realisation of 
the mathematical concept
of the classical ideal fluid obeying the Euler equation.
On the other hand, quantized vortex-lines \textit{can} 
reconnect modifying the flow topology
which, hence, is not frozen, in contrast to ideal classical fluids. 
Questions naturally arise:
is there an analogue of classical helicity in superfluid helium and other
quantum fluids? What is its physical meaning? 
And, is it conserved? 

These issues are currently debated in the context of 
the Gross-Pitaevskii equation (GPE).
In GPE theory \cite{Primer}, the vortex centerline (the vortex axis) 
is a nodal line of the wavefunction
surrounded by a thin tubular core region of depleted density. 
Since the GPE vorticity, $\bom$, is a Dirac delta function on the centerline
(where the velocity $\boldv$ diverges),
a definition of superfluid helicity based on 
Eq.~\refeq{eq:H1} requires a careful limiting procedure for $r \to 0$,
where $r$ is the radial distance to the centerline \cite{diLeoni2017}, 
and raises the question as to whether $\mathcal{H}$,
thus defined, is conserved or not.
An alternative approach is to define the
superfluid helicity using the
classical decomposition \cite{MoffattRicca1992} of 
$\mathcal{H}$ for thin vortex tubes into writhe, link and twist,
in which case superfluid helicity is zero at all times
\cite{Hanninen2016,ZuccherRicca2018,Salman2018}. 
The second approach has a subtle aspect:
the twist has an intrinsic component which relies upon the construction of
a second vortex line along the vortex centerline in order to define a ribbon; this
creates a difficulty in the context of GPE theory where there is only one 
vortex line in the core, the centerline delta function.

At the mesoscale level of description of superfluid
turbulence provided by the VFM, the definition of superfluid helicity
should not depend on the detailed nature of the vortex core, 
provided the definition has not physical or mathematical inconsistencies.
Although the GPE is a good quantitative model of gaseous Bose-Einstein
condensates, it is only a qualitative model of helium~II,
which is a liquid, not a weakly interacting Bose gas. 
A many-body quantum mechanical description of the helium vortex core
accounts for rotons \cite{Amelio2018} and a more structured vortex core 
\cite{GalliReattoRossi2014} than predicted by the GPE, revealing that the
the azimuthal velocity has
the form of a Rankine vortex with crossover from $v_{\theta} \sim r$ behaviour 
to $v_{\theta} \sim 1/r$ behaviour at $r \approx a_0$, unlike the
vortex solution of the GPE which is $v_{\theta} \sim 1/r$ for all $r \neq 0$.
The vorticity $\bom$ is therefore zero 
but in narrow tubes of approximately constant cross section $\pi a_0^2$ 
with magnitude $\omega=\kappa/(\pi a_0^2)$ and direction tangential 
to the centerline.  This result justifies the application of
the classical definition of helicity, Eq.~\refeq{eq:H1}, to helium~II.
At the mesoscale level we find (see Appendix 1):

\begin{equation}
\mathcal{H}(t)=
\kappa \; \sum_{i=1}^{N_v} \int_{_0}^{^{L_i}}\!\!\!\! 
d\xi  \; \bolds'(\xi,t)\cdot \boldv^{non}(\bolds(\xi,t),t) \; ,
\label{eq:H2}
\end{equation}

\noindent
where $\boldv^{non}(\bolds(\xi,t),t)= \boldv^{ext}(\bolds(\xi,t),t) 
+ \boldv^{far}(\bolds(\xi,t),t)$
is the {\it non-local} velocity \cite{Sherwin2012} 
induced at a point $\bolds$ along a line 
by distant vortex line elements or by external means.

In order to compare different vortex configurations,
the mesoscale helicity $\cal{H}$ must be normalized by the vortex length $L$,
because, if the helicity density 
$h(\xi)=\bolds'(\xi)\cdot \boldv^{non}(\bolds(\xi))$
is constant along the vortex lines, $\cal{H}$ would simply be proportional
to $L$.  If we divide $\mathcal{H}$ by $\kappa L$,
we find that the tangle-averaged non-local velocity contribution 
in the direction of the vorticity
is the ratio of mesoscale helicity to the quantum of circulation 
times the total vortex length, \textit{i.e.}:

\begin{equation}
< \bolds' \cdot \boldv^{non} >=\frac{\mathcal{H}}{\kappa L}\; .
\label{eq:H3}
\end{equation}

\noindent
Eq.~\refeq{eq:H3} provides a physical interpretation of 
mesoscale helicity.  We expect therefore that, 
if the vortex lines are randomly oriented in space (Vinen turbulence),
then $\mathcal{H}$ will be approximately zero
because the non-local velocity contributions at a point $\bolds$ 
on a vortex line will tend to cancel out each others. Vice versa,
if the vortex tangle is partially polarised via bundles of quasi-parallel 
vortex lines within a random tangle (Kolmogorov turbulence), 
$\mathcal{H}$ will be nonzero.
Notice that the existence of large scale flows 
is not a sufficient condition for large helical flows: for example,
two straight parallel vortex lines create a large scale
superflow but the total mesoscale helicity is zero.  

To test the interpretation of mesoscale helicity outlined in the 
previous section, we perform numerical simulations using the VFM.
In Section V we examine the mesoscale helicity $\mathcal{H}$
of simple vortex configurations. 
In Section VI we determine $\mathcal{H}$
for two distinct turbulent vortex configurations.

\section{Simple Vortex Configurations.} 

\begin{figure}
\begin{center}
\begin{minipage}{0.48\columnwidth}
  \centering
  \includegraphics[width=0.95\columnwidth]{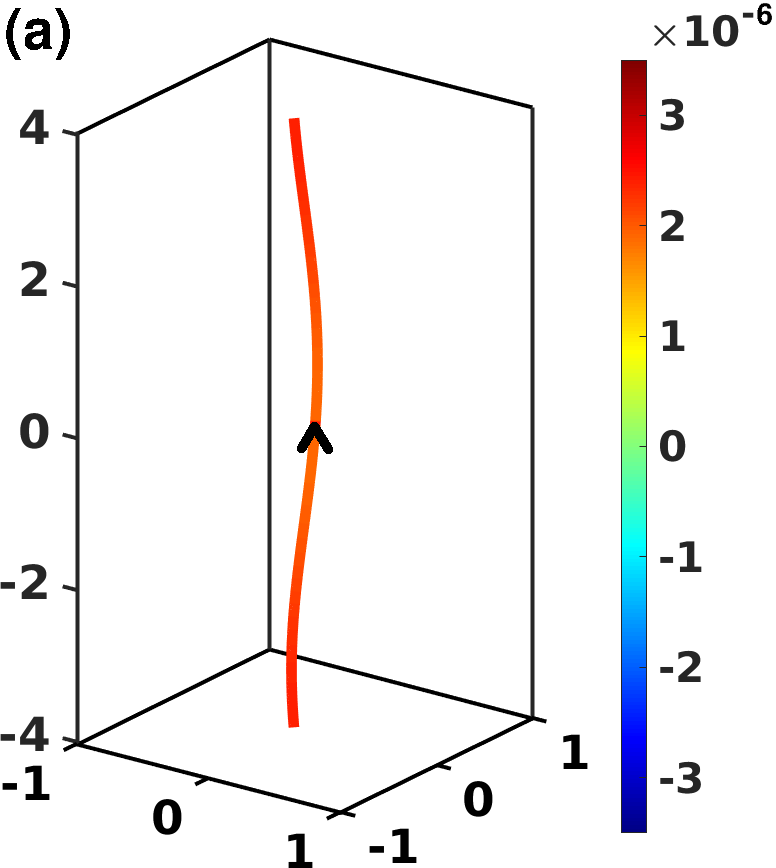}
\end{minipage}
\hspace{0.01\columnwidth}
\begin{minipage}{0.48\columnwidth}
  \centering
  \includegraphics[width=0.95\columnwidth]{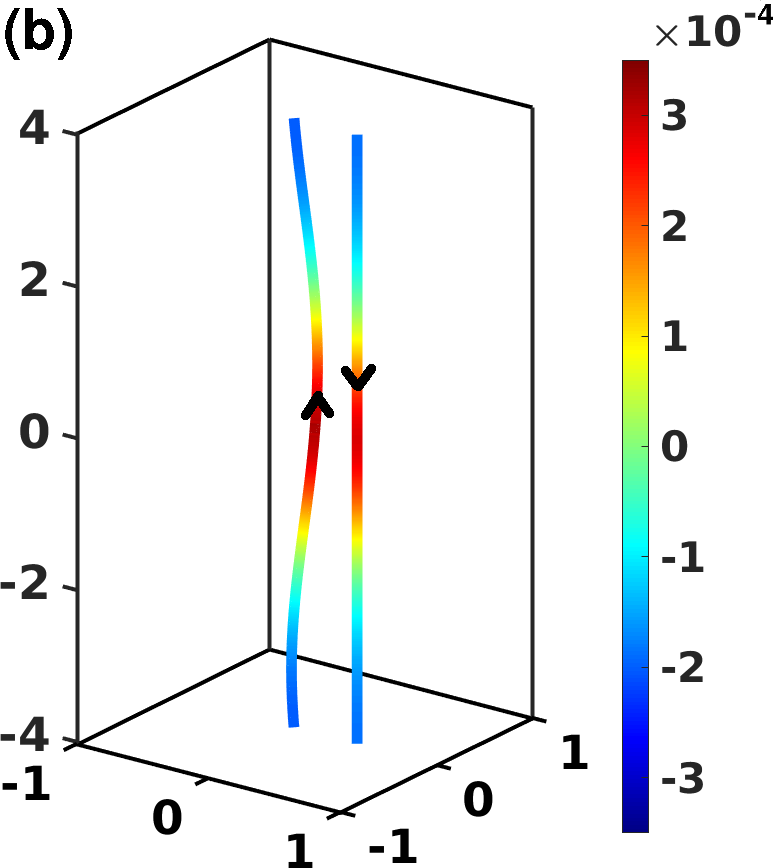}
\end{minipage}\\[5mm]
\begin{minipage}{0.48\columnwidth}
  \centering
  \includegraphics[width=0.95\columnwidth]{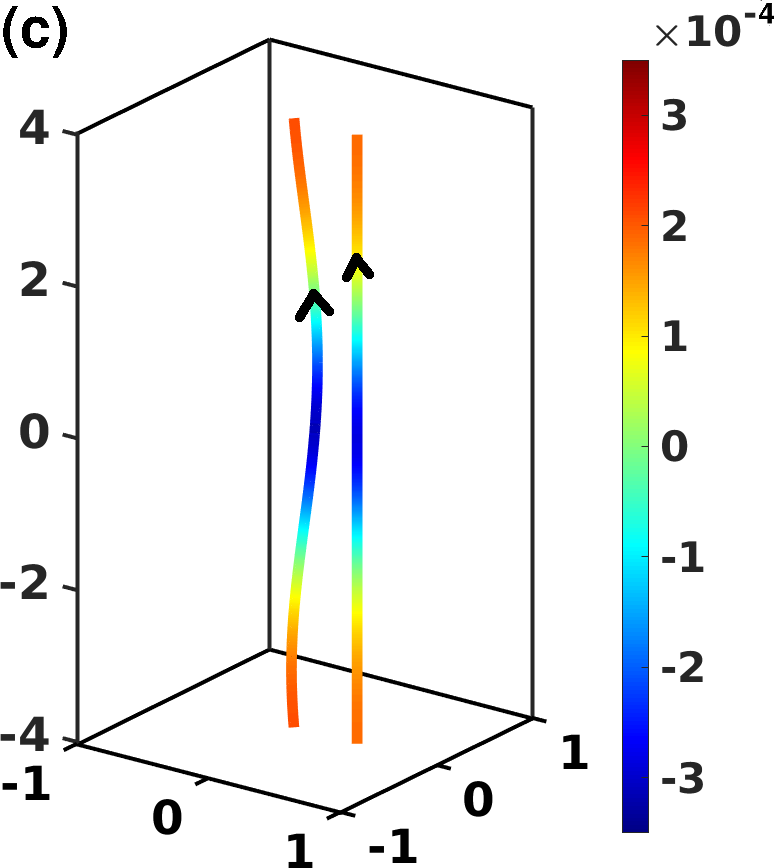}
\end{minipage}
\hspace{0.01\columnwidth}
\begin{minipage}{0.48\columnwidth}
  \centering
  \includegraphics[width=0.95\columnwidth]{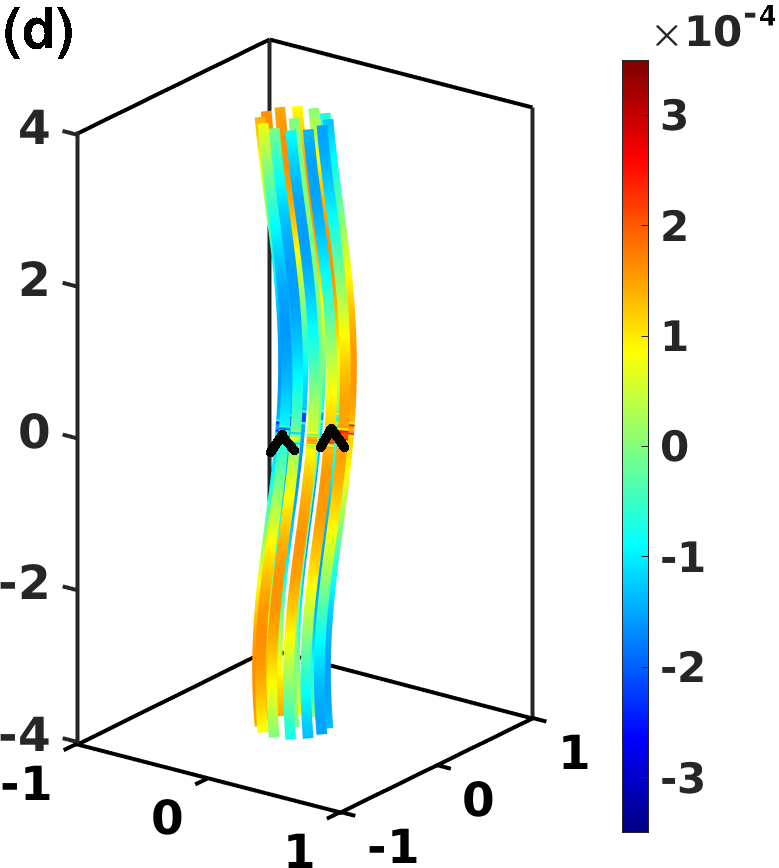}\\[1mm]
\end{minipage}
\caption{
The four simple vortex configurations (described in the main text)
used to illustrate the physical meaning of the mesoscale 
helicity $\mathcal{H}$ defined in Eq.~\refeq{eq:H2}. 
The computed values of $\mathcal{H}$ are as follows: 
(a) $\mathcal{H}/\kappa = 1.72\times 10^{-6}\;\text{cm}^2 /\text{s} $;
(b) $\mathcal{H}/\kappa = 1.96\times 10^{-6}\;\text{cm}^2 /\text{s} $;
(c) $\mathcal{H}/\kappa = 1.48\times 10^{-6}\;\text{cm}^2 /\text{s} $;
(d) $\mathcal{H}/\kappa = -1.10\times 10^{-5}\;\text{cm}^2 /\text{s} $.
Vortices are coloured according to the local value of the helicity
density $h(\xi)$. Arrows indicate the orientation of the vortices. The unit
of length shown on the axes is $10^{-1}\;\text{cm}$.}
\label{fig:simple_config}
\end{center}
\end{figure}

In order to further understand the physical meaning of the
mesoscale helicity $\mathcal{H}$ defined in Eq.~\refeq{eq:H2}, 
we numerically compute it for the following simple vortex configurations: 
(a) a vortex line perturbed by a Kelvin wave; 
(b) a vortex line perturbed by a Kelvin wave 
together with a second straight
vortex line placed alongside the first line and
oriented in the anti-parallel direction; 
(c) a vortex line perturbed by a Kelvin wave together with a second
straight vortex line placed alongside the first line and oriented
in the parallel direction; 
(d) a bundle of 13 parallel vortex lines all perturbed by Kelvin waves. 
The four vortex configurations are displayed in 
Fig.~\ref{fig:simple_config}, where vortices are coloured according to the 
value of the local mesoscale helicity density 
$h(\xi) = \bolds'(\xi)\cdot \boldv^{non}(\bolds(\xi))$.

In all configurations (a) - (d), the Kelvin wave amplitude is 
$A=8\times 10^{-3} \;\text{cm}$, its wavelength is $\lambda = 0.8 \;\text{cm}$, 
the size of the cubic computational box is 
$D=\lambda$, the spatial discretization along the vortex lines is 
$\Delta\xi=1.5\times 10^{-3}\; \text{cm}$ 
and periodic boundary conditions are employed. 
In cases (b) and (c) the distance between the perturbed vortex line
and the straight vortex line is $d_{lines} = 4\times 10^{-2}\;\text{cm}$, while 
in case (d) 12 vortex lines are arranged circularly around a central 
vortex line at distance $d_{KW} = 2\times 10^{-2}\;\text{cm}$.
As a consequence, given that $d_{lines}/D=5 \times 10^{-2}$ and 
$d_{KW}/D=2.5 \times 10^{-2}$ boundary effects can be neglected.

To interpret the resulting values of $\mathcal{H}$,
listed in the caption of Fig.~\ref{fig:simple_config},
it is useful to decompose $\mathcal{H}$ 
into {\it self-induced} and {\it interaction} contributions:
$\mathcal{H}=\mathcal{H}_{self}+\mathcal{H}_{int}$.
The first contribution, $\mathcal{H}_{self}$,
is the sum of the helicity of each vortex line arising from the
non-local velocity generated on the line by the line itself 
(implying that $\mathcal{H}_{self} = 0$ for a single straight vortex line);
the second contribution, $\mathcal{H}_{int}$,
is the helicity of the vortex lines due to the non-local velocity
field generated by the other vortices. 
The mesoscale helicities of 
configurations (a), (b) and (c) can therefore be written as
$\mathcal{H}^{(a)} = \mathcal{H}_{self}^{(a)}$,
$\mathcal{H}^{(b)} = \mathcal{H}^{(a)} + \mathcal{H}_{int}^{(b)}$ and 
$\mathcal{H}^{(c)} = \mathcal{H}^{(a)} + \mathcal{H}_{int}^{(c)}$.

We note that a Kelvin wave, arising from an instability of a
straight vortex line, has a natural orientation and rotates about
the initial unperturbed vortex in direction opposite to the superfluid
velocity.
 We find that when we add an anti-parallel
straight vortex line to the perturbed vortex
the total mesoscale helicity $\mathcal{H}/\kappa$ increases, while when
we add a parallel straight vortex line, the total mesoscale helicity
$\mathcal{H}/\kappa$ decreases.

The reason of this behaviour is two-fold. Firstly, 
$\mathcal{H}_{int}^{(c)} = - \mathcal{H}_{int}^{(b)}$
because the local helicity density $h(\xi)$ changes sign 
as the orientation of the straight
line is reversed. Secondly, $\mathcal{H}_{int}^{(b)} > 0$, 
because
in case (b) the line elements of the Kelvin wave which are closer to the
straight vortex mutually induce onto each other a positive interaction 
helicity density $h_{int}$ whose value is larger than the absolute value 
of the negative helicity density $h_{int}$
resulting from the interaction of more distant line elements: 
the overall sum $\mathcal{H}_{int}^{(b)}$ is thus positive.
More precisely, we find  
$\mathcal{H}_{int}^{(b)}/\kappa = 2.4 \times 10^{-7}\;\text{cm}^2 /\text{s}$ 
in case (b) and  
$\mathcal{H}_{int}^{(c)}/\kappa = - 2.4 \times 10^{-7}\;\text{cm}^2 /\text{s}$
in case (c), leading
to the corresponding total values reported in Fig.~\ref{fig:simple_config}. 
If we repeat the computation of $\mathcal{H}_{int}$ for 
two parallel vortex lines both perturbed by the same Kelvin wave as in (a) 
and separated by the same distance $d_{lines}$ as in (b) and (c), we obtain that
$\mathcal{H}_{int}/\kappa = -4.1 \times 10^{-7}\;\text{cm}^2 /\text{s}$.
As term of reference, if in case (a) we employ Eq. (\ref{eq:H3}) 
to estimate the magnitude
of the tangle-averaged non local velocity contribution in the 
direction of the vorticity $v^{non}$, we obtain 
$v^{non}=2.2 \times 10^{-6}\;\text{cm} /\text{s}$, which, 
in this configuration, is 
significantly smaller than the self-induced velocity along the Kelvin wave
$v^{near}=7.6 \times 10^{-4}\;\text{cm} /\text{s}$, computed using Eq. (\ref{eq:loc}) 
and employing the radius of
curvature of the Kelvin wave.

The negative values of $\mathcal{H}_{int}$ in the circumstance of 
quasi-parallel vortex lines 
is the reason why the overall mesoscale helicity  $\mathcal{H}$ 
is negative if the vortex configuration consists of a bundle of Kelvin waves
as in Fig.~\ref{fig:simple_config} (d). 
In this configuration, the $i$-th vortex line has a self-induced 
helicity $\mathcal{H}_{self}^i = \mathcal{H}^{(a)} > 0$, hence
$\displaystyle \mathcal{H}_{self} = \sum_{i=1}^N\mathcal{H}_{self}^i>0$. 
However, $\mathcal{H}_{int}$ is a large negative number 
because it is the sum of $N(N-1)/2$ interaction terms which are 
individually negative (here $N=13$).
Overall, the total mesoscale helicity 
$\mathcal{H}=\mathcal{H}_{self}+\mathcal{H}_{int}$ is negative.

We conclude that in superfluid turbulence, bundles of quasi-parallel curved vortex lines (which are
responsible for the existence of large scale flows and quasi-classical
Kolmogorov energy spectra \cite{Baggaley2012bundles})
possess negative helicity.  

\section{Turbulent Vortex Configurations}

In the following subsections we determine the mesoscale helicity
$\mathcal{H}$ of two distinct turbulent configurations which we refer to
as thermally-driven turbulence and injection-driven turbulence. 
In all numerical simulations of this section,
the computational domain is a cube of size $D=1~\rm cm$
with periodic boundary conditions. The Lagrangian
spatial discretization along the vortex lines is typically
$\Delta\xi=1.5 \times 10^{-2}~\rm cm$ to
$2.0 \times 10^{-2}~\rm cm$,
time step is $\Delta t = 4 \times 10^{-3}~\rm s$ to
$5 \times 10^{-3}~\rm s$,
and the time evolution is computed using a third order Runge-Kutta scheme.
Vortex reconnections \cite{koplik-levine-1993,serafini-etal-2017,Galantucci2019} are implemented algorithmically
as described elsewhere \cite{Schwarz1988,Baggaley2012recon} and 
the Biot-Savart integral Eq.~\refeq{eq:BS1} is computed via a 
tree approximation \cite{baggaley-barenghi-2011}
in order to decrease the computational cost of the calculations. 
All simulations start with a few seeding vortex lines which quickly 
multiply, until, after an initial transient, a statistical steady-state 
of turbulence is achieved in which the vortex line density $\mathcal{L}$
(or, equivalently, the vortex length $L=\mathcal{L}D^3$)
fluctuates around a mean value independent of the initial
condition, as shown by the blue curves in Figures~\ref{fig:CF_HL} and \ref{fig:inj_HL}, 
for the two experiments. In this steady-state, a balance is reached between
forcing and dissipation. Forcing occurs via the Donnelly-Glaberson instability 
\cite{Tsubota2003} in thermally-driven turbulence
where $T>0$, 
and via vortex ring injection in injection-driven turbulence; 
dissipation takes place via the friction in thermally-driven turbulence 
and via the finite discretization on the vortex lines which models the damping of the shortest Kelvin 
waves by phonon emission in injection-driven turbulence 
where $T = 0$.

\subsection{Thermally-driven turbulence}

In this numerical simulation, we assume a uniform thermal
counterflow velocity $\boldv_{ns} = \boldv_{n} - \boldv^{ext}$ in Eq.~\refeq{eq:Schwarz},
and simulate Vinen turbulence driven thermally by a small heat 
flux in a large channel \cite{Sherwin2012} at nonzero temperature. We choose
temperature $T=1.9~\rm K$ (corresponding to $\alpha=0.206$
and $\alpha'=0.009$) and $\vert \boldv_{ns} \vert =0.08~\rm cm/s$.

\begin{figure}[htbp]
\includegraphics[width=0.99\linewidth]{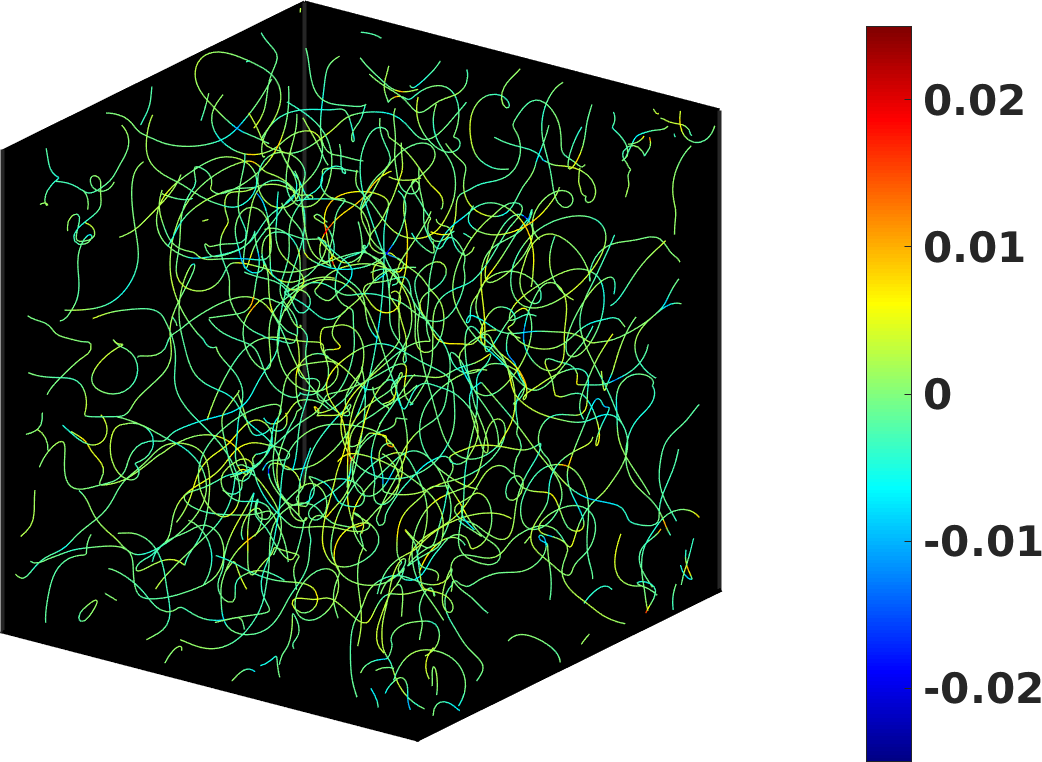}
\caption{
Thermally-driven turbulence. Snapshot of turbulent vortex tangles  
driven by uniform normal fluid (Vinen regime) at $t=2400 \; \rm{s}$. 
The vortex lines are colour-coded according to the 
local mesoscale helicity density $ h(\xi) = \bolds'(\xi)\cdot \boldv^{non}(\bolds(\xi)) $.}
\label{fig:CF_snap}
\end{figure}
\begin{figure}[htbp]
\includegraphics[width=0.99\linewidth]{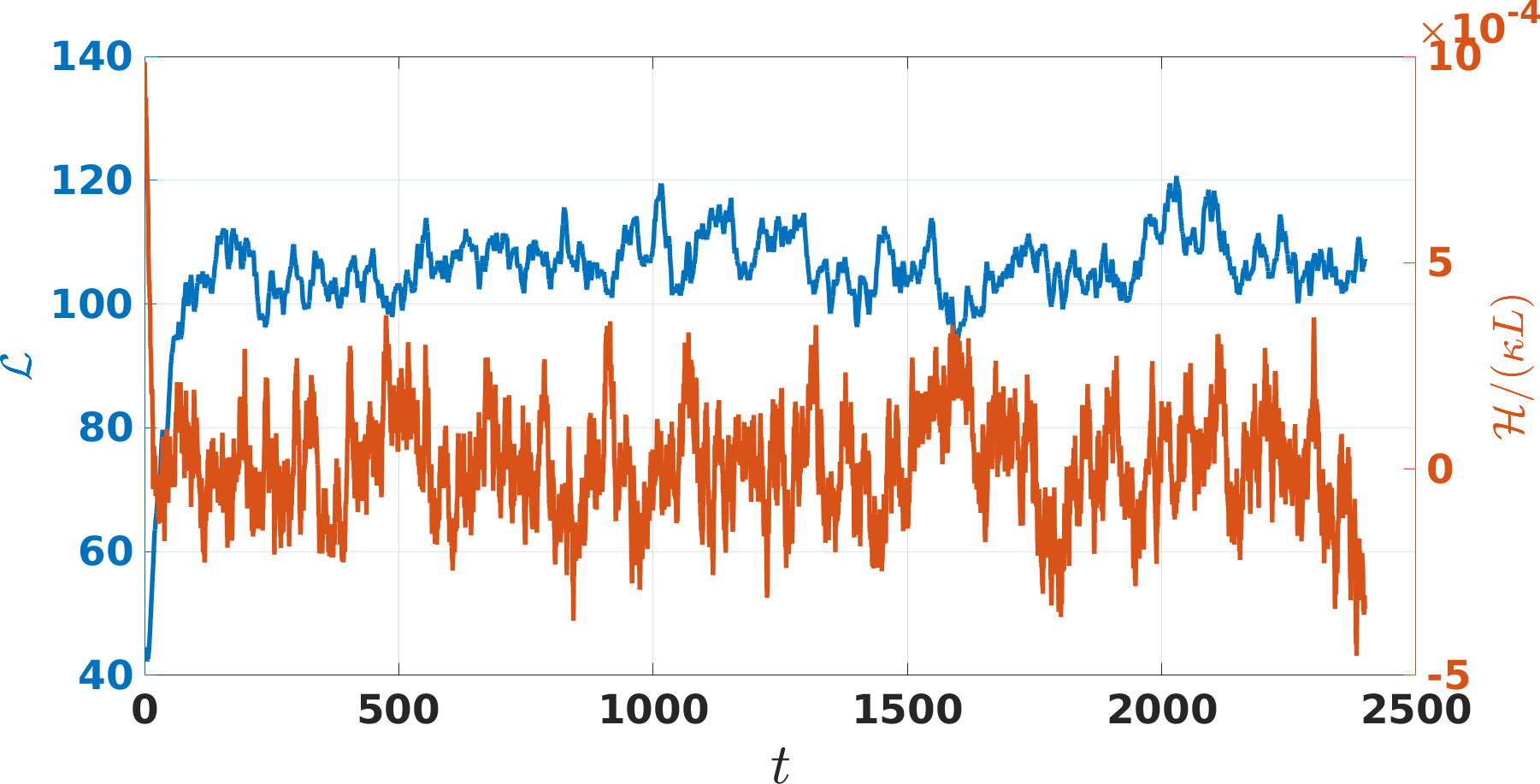}
\caption{Thermally-driven turbulence.
Temporal evolution of vortex line density $\mathcal{L}(t)$ ($\text{cm}^{-2}$, 
blue curve) and
mesoscale helicity ${\cal{H}}$ divided by $\kappa L$ ($\text{cm/s}$, red curve) for thermal counterflow. 
Time $t$ is indicated in seconds.}
\label{fig:CF_HL}
\end{figure}

In Fig.~\ref{fig:CF_snap} we report a snapshot of the vortex tangle
after the initial transient, where vortex lines are colour-coded 
according to the local value of the mesoscale helicity density
$ h(\xi) = \bolds'(\xi)\cdot \boldv^{non}(\bolds(\xi)) $.
It emerges that in the Vinen regime
the mesoscale helicity density $ h(\xi)$ is essentially zero
everywhere.
This reflects in the temporal evolution
of the integrated mesoscale helicity ${\cal{H}}$ divided by $\kappa L$ (reported in Fig.~\ref{fig:CF_HL}), 
where ${\cal{H}}$ performs small oscillations around zero 
when the statistically steady-state is achieved.

\subsection{Injection-driven turbulence}

\begin{figure}[htbp]
\includegraphics[width=0.99\linewidth]{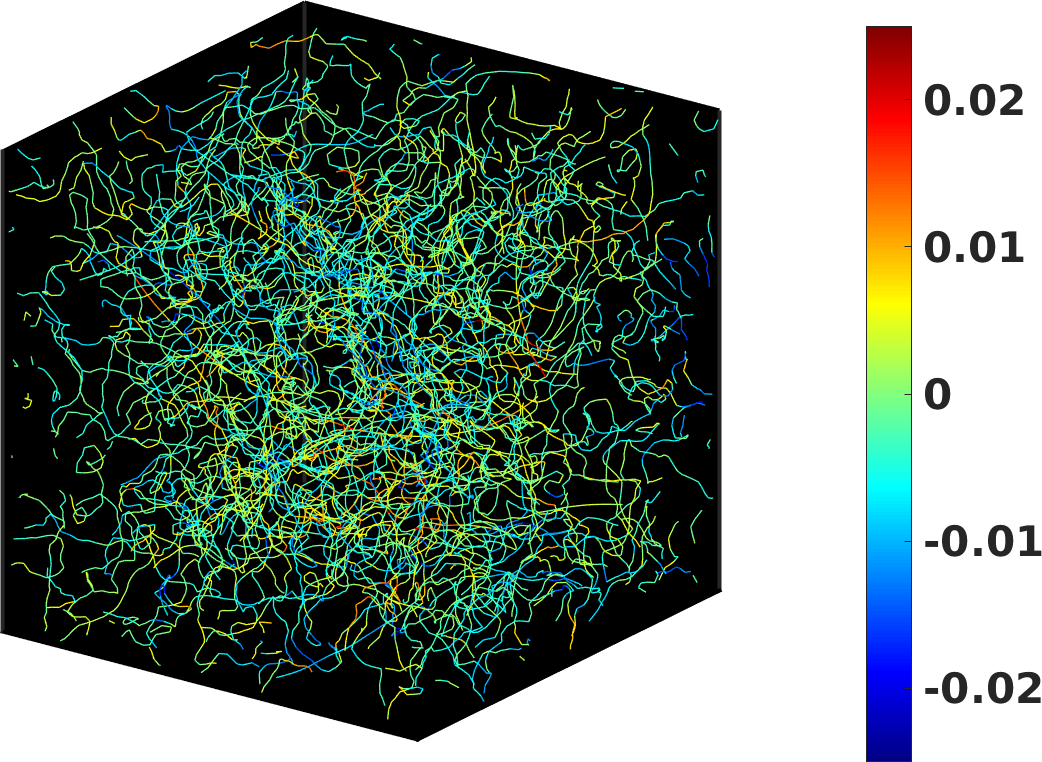}
\caption{
Injection-driven turbulence. Snapshot of turbulent vortex tangles  
driven by constant vortex ring injection at zero temperature at $t=175 \; \rm{s}$. 
The vortex lines are colour-coded according to the 
local mesoscale helicity density $ h(\xi) = \bolds'(\xi)\cdot \boldv^{non}(\bolds(\xi)) $.}
\label{fig:inj_snap}
\end{figure}

\begin{figure}[htbp]
\includegraphics[width=0.99\linewidth]{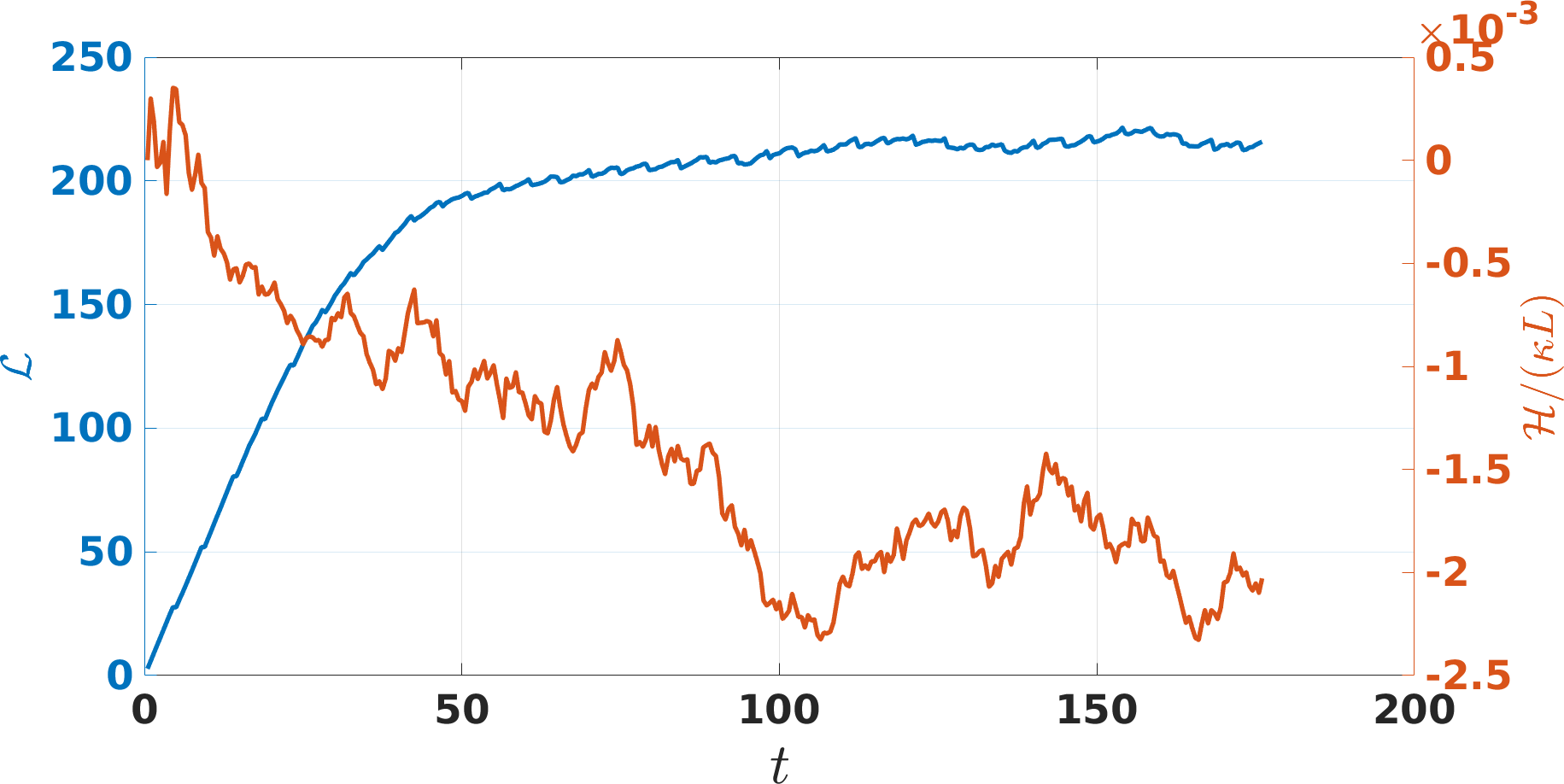}
\caption{
Injection-driven turbulence.
Temporal evolution of vortex line density $\mathcal{L}(t)$ ($\text{cm}^{-2}$, blue curve) and
mesoscale helicity ${\cal{H}}$ divided by $\kappa L$ ($\text{cm/s}$, red curve) 
for superfluid turbulence driven by constant vortex ring injection at $T=0$ . 
Time $t$ is indicated in seconds.
}
\label{fig:inj_HL}
\end{figure}

In this numerical experiment we generate turbulence at $T=0$ (no normal fluid)
by injecting vortex rings of radius $R=D/2$ 
at random positions and with random orientation at the constant rate
$\dot{\cal{L}}=5.6~ \text{cm}^{-2}/\text{s}$ (a similar set-up was used in the 
laboratory by Walmsley and Golov \cite{Walmsley2008}).
The snapshot of the vortex tangle in the statistically steady-state colored 
according to the local mesoscale helicity density $h(\xi)$ 
(Fig.~\ref{fig:inj_snap})
shows that locally $\vert h(\xi) \vert$ 
achieves significantly larger values if compared
to Vinen turbulence (Fig.~\ref{fig:CF_snap}). 
Furthermore, regions with negative $h(\xi)$ 
seem to prevail over
regions with positive $h(\xi)$. 
This visual suggestion is confirmed by the temporal
evolution of $\mathcal{H}/(\kappa L)$ reported in Fig.~\ref{fig:inj_HL} which shows
that, after a transient, $\mathcal{H}/(\kappa L)$ is negative and its magnitude is larger 
than the one reported for Vinen turbulence (shown in Fig.~\ref{fig:CF_HL}).


\section{Discussion}

\begin{figure}
\centering
\includegraphics[width=0.75\columnwidth]{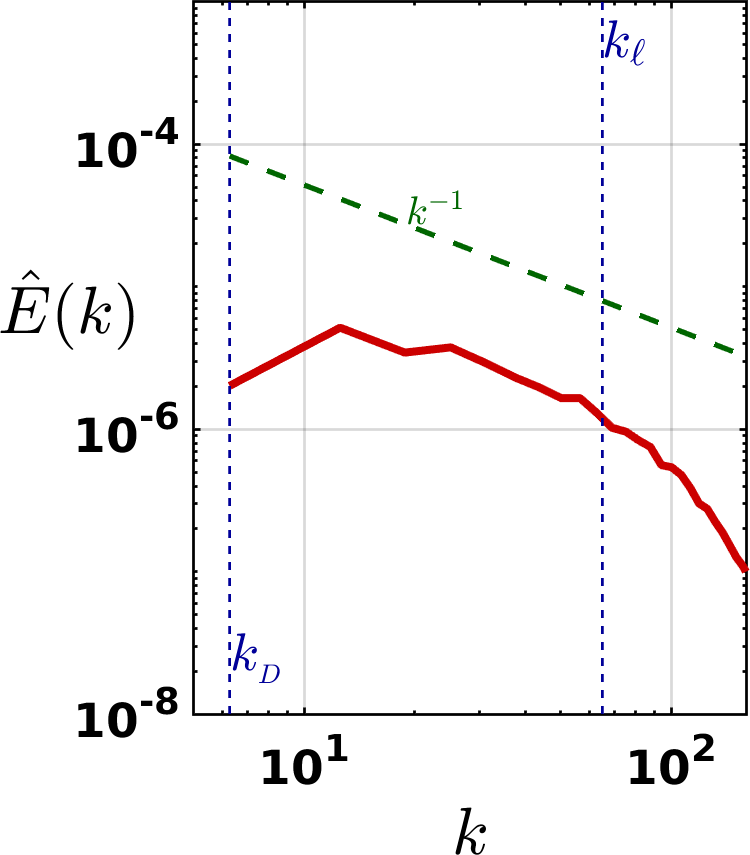}
\caption{
Thermally-driven turbulence. Counterflow superfluid 
turbulence at $T=1.9~\rm K$:
energy spectrum ${\hat{E}}$ vs wavenumber $k$ at $t=2400 \; \rm{s}$; the dashed green line
indicates the typical $k^{-1}$ isolated vortex scaling, while blue 
vertical lines indicate wavenumbers corresponding to the size of the 
box $D$ and the average intervortex spacing $\ell$. }
\label{fig:CF_spectra}
\end{figure}

We interpret the observed behaviours of the magnitude 
of $\mathcal{H}/(\kappa L)$ in thermally-driven
and injection-driven turbulence as follows.
In thermally-driven (Vinen) superfluid turbulence 
the vortex tangle configuration is a random 
arrangement of vortex loops and lacks
large-scale flows. This is confirmed by the kinetic energy spectrum  
$\hat{E}(k)$ which we report in Fig.~ \ref{fig:CF_spectra}, which shows 
the characteristic $\hat{E}(k) \sim k^{-1}$
behaviour of an isolated vortex at $k\approx k_\ell$, where $k_\ell = 2\pi /\ell$ and $\ell=1/\sqrt{\mathcal{L}}$
is the average inter-vortex spacing. The absence of large scale 
flows apparent in figure ($\hat{E}(k)$ decreases for
$k \to k_D$) implies that the non-local 
velocity contributions $\boldv^{non}$ in Eq.~\refeq{eq:H2} 
are small (as contributions from
distinct vortices tend to cancel out given the random configuration), 
leading to an overall
smaller value of  $\mathcal{H}/(\kappa L)$ observed in Fig.~\ref{fig:CF_HL} 
with respect to the value of $\mathcal{H}/(\kappa L)$ calculated for 
injection-driven turbulence (Fig.~\ref{fig:inj_HL}).

\begin{figure}
\centering
\includegraphics[width=0.75\linewidth]{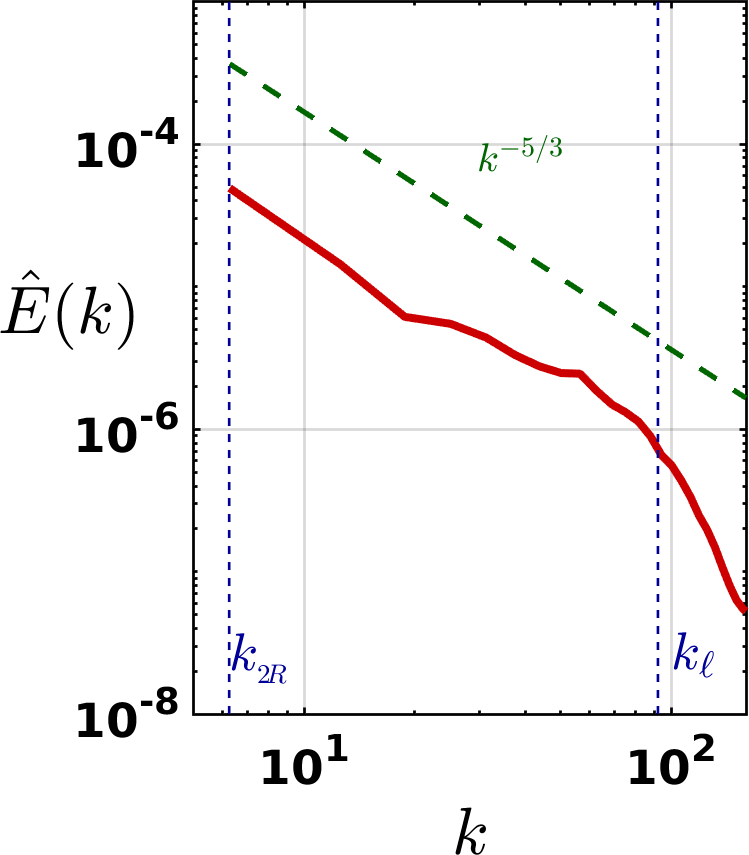}
\caption{
Injection-driven turbulence. Superfluid turbulence driven by injection of vortex rings at $T=0$:
energy spectrum ${\hat{E}}$ vs wavenumber $k$ at $t=175 \; \rm{s}$; the dashed green line
shows the slope of the classical Kolmogorov $k^{-5/3}$ scaling, while blue vertical lines
indicate wavenumbers corresponding to the diameter $2R$ of the injected vortex rings and the average intervortex spacing $\ell$.}
\label{fig:inj_spectra}
\end{figure}

If the superfluid turbulence is driven by vortex ring injection instead,
the energy spectra $\hat{E}(k)$ reveal a quasi-classical Kolmogorov behaviour,
with $\hat{E}(k) \sim k^{-5/3}$ for $k \ll k_\ell$, 
as shown if Fig.~\ref{fig:inj_spectra}.
This property implies that most of the energy is contained in
large-scale eddies ($\hat{E}(k)$ increases for $k \to k_D$). 
Thus, $\boldv^{non}$ is significantly larger than in counterflow turbulence, 
accounting, see Eq.~\refeq{eq:H3}, for the observed larger magnitude 
of ${\mathcal{H}/(\kappa L)}$ when compared to thermally-driven turbulence. 
Furthermore, on the basis of the results presented in 
Section~V, we ascribe the negative value of 
${\mathcal{H}/(\kappa L)}$ in injection-driven turbulence
to the presence in the flow of vortex-line bundles which are 
responsible for the observed
Kolmogorov spectrum \cite{BaggaleyLaurie2012,Baggaley2012bundles}.
For the sake of completeness, in Appendix 3 we report the temporal
evolution of mesoscale helicity $\mathcal{H}$ in a
numerical simulation of mechanically-driven superfluid 
turbulence\cite{Bauer1997,Sherwin2012}.

The interpretation of the mesoscale
helicity as a measure of the non-local contribution to the motion of the
vortex lines, Eq.~(\ref{eq:H3}), is confirmed by expressing ${\cal H}$
in units of the characteristic self-induced 
velocity of the vortex lines in the tangle, which can be estimated from Eq.~(\ref{eq:loc})
using $R \approx 1/\sqrt{\cal L}$.
For the thermally-driven turbulence shown in
Fig.~\ref{fig:CF_snap}, we obtain $v^{non}/v^{near} \approx 0$,
as $\cal H$ fluctuates around zero; the largest fluctuations suggest that 
$v^{non}/v^{near}$ is at most $0.8 \%$. This is consistent with the
early finding of Schwarz\cite{Schwarz1988} that the self-induced velocity
alone is enough to generate the observed intensity of thermally-induced
turbulence.
For the injection-driven turbulence shown in
Fig.~(\ref{fig:inj_snap}), we obtain
$v^{non}/v^{near} \approx 11 \%$, a significant contribution to the
total velocity, as expected.  
The model of mechanically-driven turbulence described in
Appendix 3 (Fig.~\ref{fig:ABC_snap}) 
yields $v^{non}/v^{near} \approx 1.3$, again as expected as this flow
has the Beltrami property of maximal helicity (cf. Appendix 3 for further
details).

A final consideration concerns vortex reconnections. 
The turbulence simulations involve thousands of reconnections. 
In the VFM, the standard reconnection procedure 
\cite{Baggaley2012recon} reduces the vortex length  
(as proxy of energy) to model acoustic 
losses revealed by more microscopic 
GPE simulations \cite{Leadbeater2001,Zuccher2012}. We have therefore
carefully monitored $\mathcal{H}$  before, during and after an
individual vortex reconnection at zero temperature (no friction).
We have found (details are in Appendix 2)
that $\mathcal{H}$ is initially conserved,
consistently with theoretical \cite{Laing2015}, experimental 
\cite{scheeler-etal-2014} and numerical \cite{Zuccher2017} 
studies showing that reconnections do not \textit{immediately} 
affect the 
{\it centerline} helicity (helicity of vortex tubes without twist 
contribution). However, as the reconnection cusp relaxes and
Kelvin wavepackets are released (a feature
also observed in experiments \cite{Fonda2014}), 
the relative proportions of near and far field velocity contributions
change, resulting in an overall jump $\Delta \mathcal{H}$.
In a turbulent flow, these jumps (being either positive or 
negative) cancel out in the statistical steady state. It is therefore
unlikely that the difference in $\mathcal{H}$ between
Kolmogorov and Vinen regimes is caused by the reconnection
algorithm, more so since the reconnection rate is larger for Vinen
than Kolmogorov turbulence \cite{Sherwin2012} (polarised vortex lines
suffer less reconnections).

\section{Conclusions}

We have shown that the classical definition of helicity
can be naturally generalised to the superfluid mesoscale context of the VFM 
without inconsistency with the microscopic
nature of the helium vortex core.
We have also shown that our mesoscale definition of helicity $\cal{H}$, 
Eq.~(\ref{eq:H2}), has the remarkable
property of quantifying the far-field velocity contributions,
$v^{non}$, which are 
induced at a point along a vortex line by other vortex lines 
(or by elements on the same vortex line which are sufficiently far away),
see Eq.~(\ref{eq:H3}); 
effectively $\cal H$ measures
the non-local contribution to the vortex motion,
which in a random vortex tangle is dominated by the locally-induced
velocity, $v^{near}$.

Our numerical experiments show that in Vinen-type
turbulence these non-local contributions tend to cancel out
(corresponding to vanishing ${\mathcal{H}/(\kappa L)}$
and $v^{non}/v^{near}$), 
whereas in quasi-classical, Kolmogorov-type 
turbulence they add up (corresponding to nonzero values of 
${\mathcal{H}/(\kappa L)}$ and $v^{non}/v^{near}$),
as the tangle becomes polarised and vortex bundles generate large scale flows. 
Moreover, our numerical simulations with simple vortex structures
explain why in Kolmogorov quasi-classical 
superfluid turbulence, mesoscopic helicity $\mathcal{H}$ is negative, 
as a result of the presence of such vortex bundles,
although larger-scale simulations are needed to rule out the role of
any inhomogeneity and anisotropy of the vortex tangle
and check how helicity scales with
the tangle's size. This helical 
property of quasi-classical superfluid turbulence is a direct consequence
of the concentration of superfluid vorticity in one-dimensional vortex lines: 
quasi-classical turbulence in superfluids is in fact directly linked to the existence of quasi-parallel vortex bundles which carry 
negative helicity.  

It is also worth emphasising that the presence of non-zero superfluid helicity 
in turbulence generated by vortex ring injection at $T=0$
demonstrates that helicity is not simply injected in the superfluid flow by
the normal fluid.

At the moment the superfluid helicity which we  have defined can only
be determined in numerical simulations, where, as we have seen, 
it is a useful monitor of vortex interactions.
It is natural to ask what is
the outlook for experimental measurements of helicity.
Recent experiments on helicity in classical fluid dynamics
\cite{Scheeler2017} show that very slender vortex tubes are
required for the experimental visualisation and the interpretation of helicity.
Since there is no physical system where vorticity is
more concentrated in thin tubes than superfluid helium, the
motivation to measure helicity in superfluid helium is clear.
A direct laboratory measurement of helicity at cryogenics temperature
is not as far-fetched as it may seem.
For example,  Kelvin waves
following superfluid vortex reconnections have been observed in
the laboratory \cite{Fonda2014}. In order to reconstruct the evolution of
the mesoscale helicity, what is further needed is a 3D image of the vortex
shape, which is potentially
feasible in liquid helium as it is done at room temperature.



\bigskip
\begin{acknowledgments}
\centerline{\bf Acknowledgments}
\bigskip

We acknowledge the support of  EPSRC grant EP/R005192/1.
\end{acknowledgments}
\bigskip

\appendix {\centerline{\bf Appendix 1: Derivation of Eq.~(\ref{eq:H2})}}
\bigskip

\begin{figure}
\centering
\includegraphics[width=0.3\linewidth]{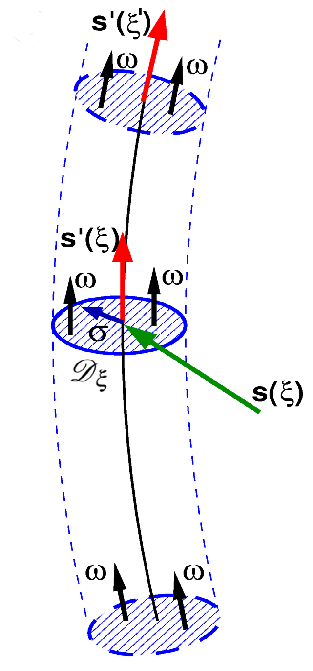}
\caption{
Schematic vortex tube
with cross-sections $\mathcal{D}_{\xi}$ of area $\pi a_0^2$.
The solid (black) line $\bolds(\xi,t)$ is the centerline.
The unit tangent vectors $\bolds'$ at $\bolds$ are the (red)
arrows. The (blue) vector $\bos$ is the
position vector on each cross-section.
Within the vortex tube, the vorticity $\bom$ is constant.
}
\label{fig:A1}
\end{figure}

\noindent
The Rankine-like velocity field observed in the many-body calculations
of the helium vortex \cite{GalliReattoRossi2014,Amelio2018}
implies that the vorticity $\bom$ is zero everywhere but in a narrow tube
of approximately constant cross sectional area $\pi a_0^2$
along the vortex line (see Figure~\ref{fig:A1})
where its magnitude is $\omega=\kappa/(\pi a_0^2)$
and its direction is tangential to the centerline. At time $t$,
the vortex configuration consists of a collection of vortices of length
$L_i$ ($i=1, \cdots N_v$), and Eq.~(1) becomes

\begin{align*}
\displaystyle
\mathcal{H}(t)&=
\sum_{i=1}^{N_v} \int_{_0}^{^{L_i}}\!\!\!\! d\xi \!\!
\int_{_{\mathcal{D}_{\xi}}} \!\!\!\!\! \bom (\xi, \bos) \cdot \boldv(\xi, \bos)
\;d^2\! \bos =  \nonumber
\\
&=
\frac{\kappa}{\pi a_0^2} \; \sum_{i=1}^{N_v} \int_{_0}^{^{L_i}}\!\!\!\! d\xi 
~\!\!\int_{_{\mathcal{D}_{\xi}}} \!\!\!\!\!
\bolds'(\xi)\cdot \boldv(\xi, \bos) d^2\!\bos,
\end{align*}

where

\begin{equation*} 
\boldv(\xi,\bos)=\boldv^{ext}(\xi, \bos)
+ \boldv^{near}(\xi, \bos) + \boldv^{far}(\xi, \bos).
\end{equation*}

Because of the large separation of length scales
between $D$, $\ell$ and $a_0$,
both the external superflow
and the superfluid velocity field induced by all the other
vortices are constant
on the disc $\mathcal{D}_{\xi}$
and can be evaluated at $\bolds(\xi)$.
In addition, the typical radius of curvature, $R_c$, is always much
larger than the vortex core radius: $R_c\sim 10^5 a_0$.
The neighbourhood $\vert \xi' -\xi \vert$ $< \delta$
on the vortex line near a point $\bolds(\xi)$
is thus effectively straight and perpendicular to
the disc $\mathcal{D}_{\xi}$ at distances $\delta$ such that $a_0 \ll \delta \ll R_c$
(range which exists given the huge scale separation characterising the system).
At such distances, the superfluid velocity $\boldv^{near}(\xi, \bos)$ induced
by the closest vortex line elements on $\mathcal{D}_{\xi}$ is perpendicular
to the unit tangent vector $\bolds'(\xi)$, yielding zero contribution
to $\mathcal{H}$.
Concerning the line element centered in $\bolds(\xi,t)$,
the only non-zero contribution to $\mathcal{H}$ which arises from the vortex
line going through $\bolds(\xi,t)$ is the contribution
of elements of that line which are
sufficiently distant from $\mathcal{D}_{\xi}$,  where the induced velocity
is constant and can be evaluated at $\bolds(\xi)$.
$\mathcal{H}$ thus reduces to Eq.~(6) in the main manuscript:

\begin{align*}
\displaystyle
\mathcal{H}(t) & = 
\kappa \; \sum_{i=1}^{N_v} \int_{_0}^{^{L_i}}\!\!\!\! d\xi  \; \bolds'(\xi)\cdot \boldv^{non}(\bolds(\xi))\; \; ,
\end{align*}

\noindent
where
\begin{equation*}
\boldv^{non}(\bolds(\xi))= \boldv^{ext}(\bolds(\xi)) 
+ \boldv^{far}(\bolds(\xi)),
\end{equation*}

\noindent
is the superfluid velocity at a point $\bolds(\xi)$ of a vortex
line induced by distant vortex line elements on that line
and other lines ($\boldv^{far}$),
added to the external potential superflow $\boldv^{ext}$.

\bigskip

\appendix{\centerline{\bf Appendix 2: Helicity  during reconnection}}
\bigskip

\begin{figure}
\includegraphics[width=0.95\linewidth]{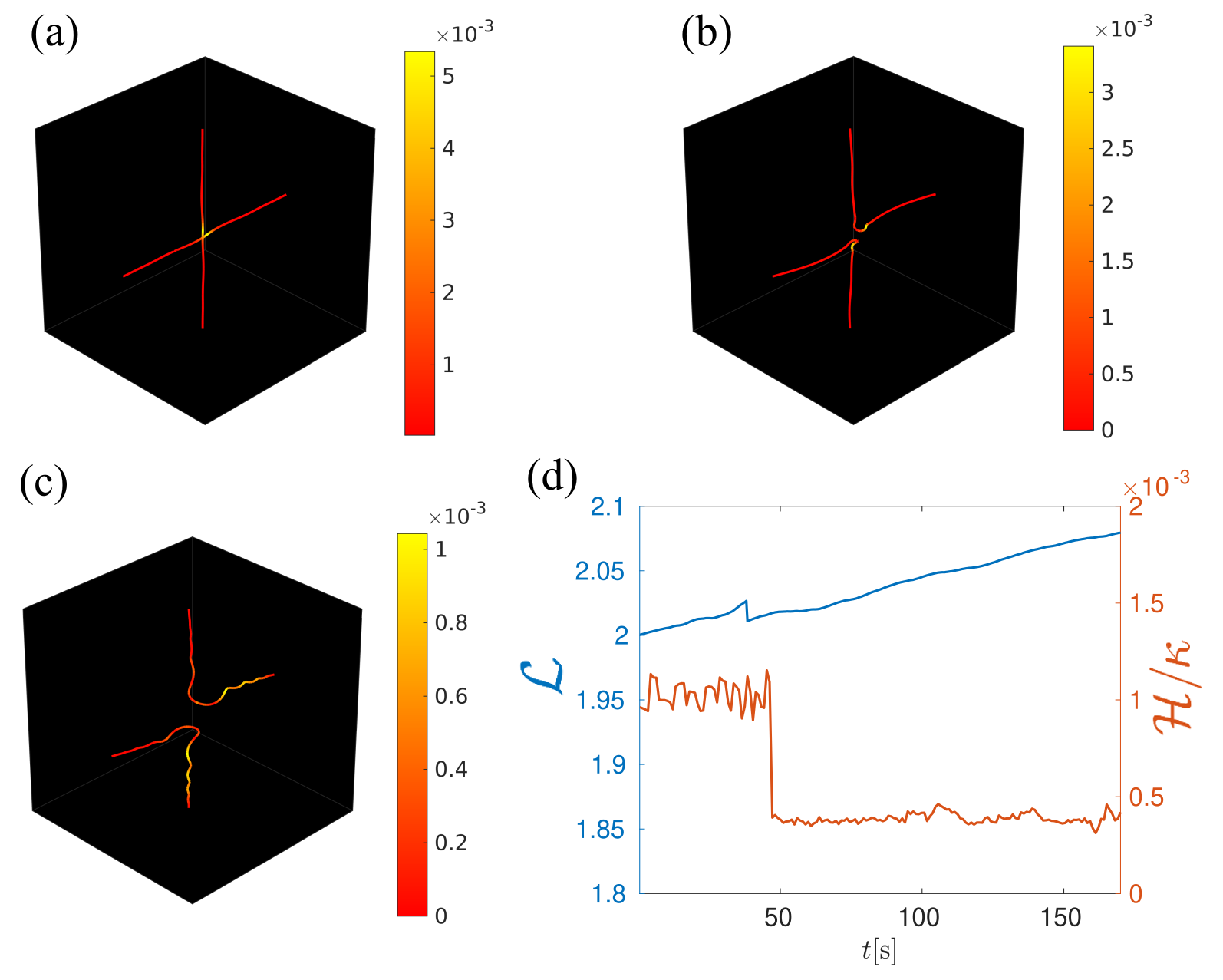}
\caption{
(a)-(c) Snapshots of two initially
orthogonal vortices undergoing a reconnection at $T=0$.
Vortex lines are colour-coded according
to the local value of the magnitude of the mesoscale helicity density
$\vert \bolds'(\xi)\cdot \boldv_{far}(\bolds(\xi)) \vert $.
(d) Evolution of the vortex line density $\mathcal{L}$ in $\text{cm}^{-2}$
 (blue)
and volume integrated mesoscale helicity $\mathcal{H}/\kappa$ 
in $\text{cm}^{2}/\text{s}$ (red).
}
\label{fig:A2}
\end{figure}

\noindent
Figure~\ref{fig:A2} illustrates a single reconnection
of two initially orthogonal vortex lines at zero temperature
simulated using the VFM in a periodic domain.
The reconnection occurs at $t\approx 48~$s, where the total length
suddenly decreases due to the reconnection algorithm (on a longer
time, the vortex lines are slightly stretched).
There is a small time delay between the reconnection and the
resulting jump $\Delta \mathcal{H}$ of mesoscale helicity caused
by the relaxing vortex cusp (notice the Kelvin waves which move
away); this jump represents the changed proportion of near and
far field velocity contributions. Before and after the
reconnection, the mesoscale helicity is constant (there is no friction,
unlike the simulations presented in the main paper).

\bigskip

\appendix{\centerline{\bf Appendix 3: Mechanically-driven turbulence}}
\bigskip

In this appendix, for the sake of completeness given previous studies
reported in the literature \cite{Bauer1997,Sherwin2012}, we perform a 
third superfluid turbulence numerical simulation.
In particular, we consider superfluid helium 
at nonzero temperature driven mechanically, \textit{e.g.} by propellers,
which induce turbulence in the normal fluid. In this case, 
we model the coherent regions of intense normal fluid 
vorticity typical of classical turbulence by imposing a steady
ABC normal fluid flow \cite{Bauer1997,Sherwin2012}
$\boldv_n=(v_{nx},v_{ny},v_{nz})$:

\begin{eqnarray}
v_{nx}&=& A \sin{(kz)}+C\cos{(ky)},\\
v_{ny}&=& B \sin{(kx)}+A\cos{(kz)},\\
v_{nz}&=& C \sin{(ky)}+B\cos{(kx)}
\end{eqnarray}

\noindent
with $A=B=C=0.03~\rm cm/s$, $k=2 \pi~\rm cm^{-1}$, and
temperature $T=1.9~\rm K$. The scale of the forcing is the
size the computational box $D$.
All numerical parameters of the simulation
(box-size, vortex-line spatial discretization and time-step)
coincide with the parameters illustrated in Section VI.

\begin{figure}[htbp]
\includegraphics[width=0.99\linewidth]{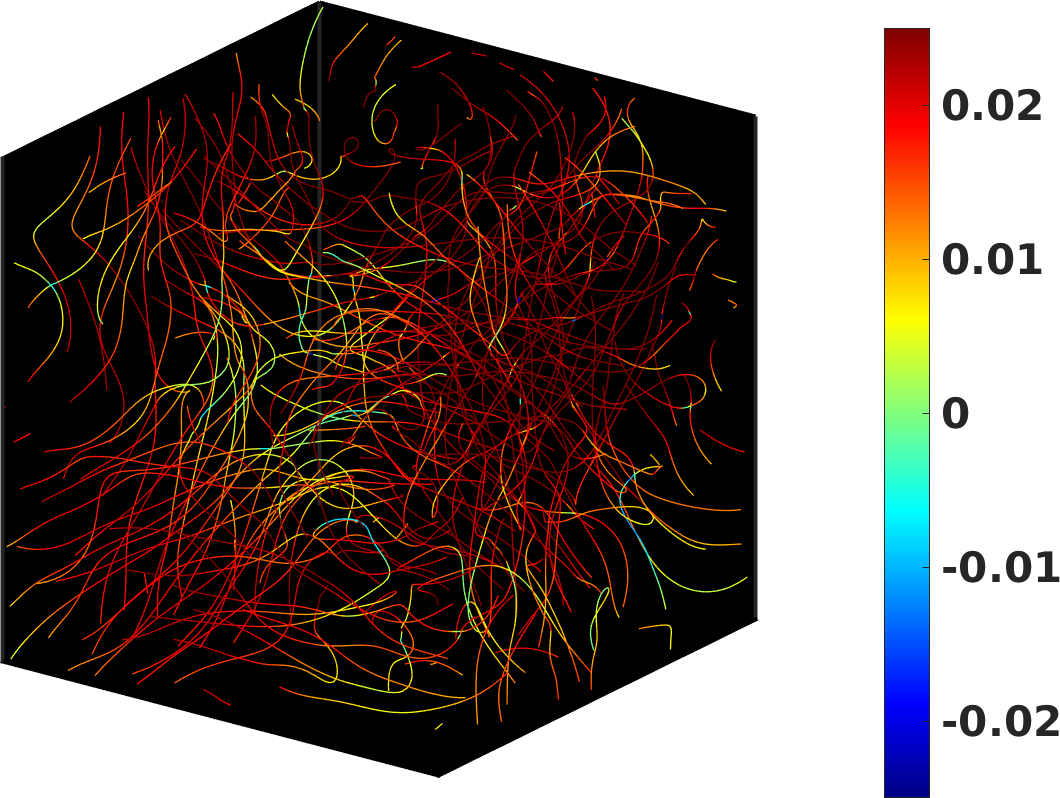}
\caption{
Mechanically-driven turbulence.
Snapshot of turbulent vortex tangle
driven by ABC normal flow at $t=2700 \; \rm{s}$. 
The vortex lines are colour-coded according to the 
local mesoscale helicity density $ h(\xi) = \bolds'(\xi)\cdot \boldv^{non}(\bolds(\xi)) $.}
\label{fig:ABC_snap}
\end{figure}

\begin{figure}[htbp]
\includegraphics[width=0.99\linewidth]{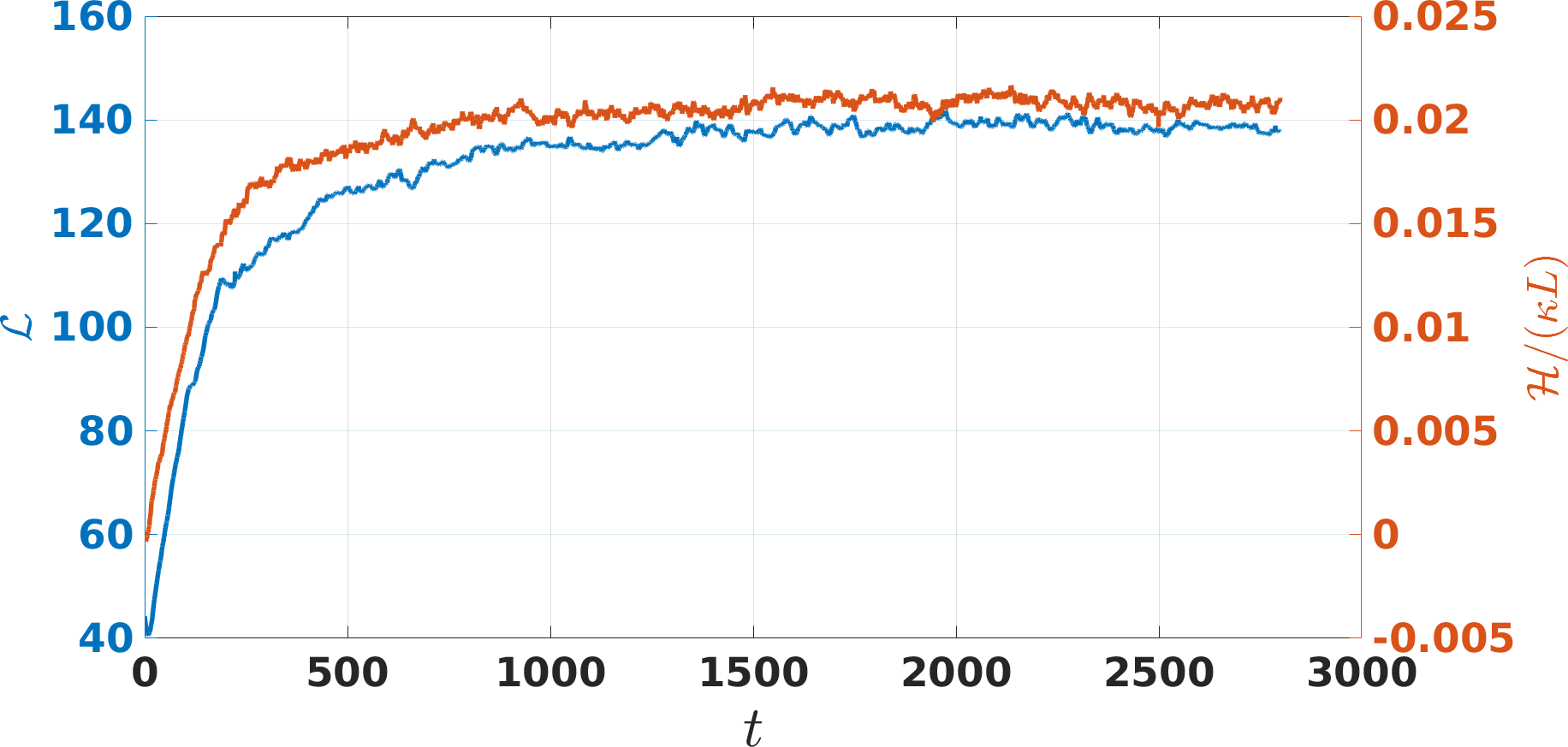}
\caption{
Mechanically-driven turbulence.
Temporal evolution of vortex line density $\mathcal{L}(t)$ ($\text{cm}^{-2}$, blue curve) and
mesoscale helicity ${\cal{H}}$ divided by $\kappa L$ ($\text{cm/s}$, red curve) for 
superfluid turbulence driven by ABC normal flow. Time $t$ is indicated in seconds.
}
\label{fig:ABC_HL}
\end{figure}

In Fig.~\ref{fig:ABC_snap} we report a helicity density colored snapshot of the vortex 
tangle in the steady-state regime: it shows regions of large magnitude of positive helicity density $h(\xi)$,
if compared to snapshots of the vortex tangle corresponding to thermally-driven and injection-driven
turbulence (Fig.~\ref{fig:CF_snap} and Fig.~\ref{fig:inj_snap}, respectively). This 
is reflected in the temporal evolution of $\mathcal{H}/(\kappa L)$ 
(shown in Fig.~\ref{fig:ABC_HL}) which in the statistically steady-state 
has a large positive value, one order of magnitude larger than the value of $\mathcal{H}/(\kappa L)$
observed in injection generated superfluid turbulence (Fig.~\ref{fig:inj_HL}).


The fact that $\mathcal{H}/(\kappa L)$ in mechanically-driven turbulence
is larger than in thermally-driven and injection-driven
turbulence is expected:
the ABC flow has the Beltrami property of maximal helicity 
\textit{i.e.} velocity and vorticity of the normal fluid are locally aligned.
As the imposed normal fluid velocity field, 
which forces the superfluid turbulence,
is stationary, superfluid vortex lines tend to align
with the normal fluid vorticity \cite{Bauer1997} boosting the
growth rate of helical Kelvin waves arising from the Glaberson-Donnelly 
instability. It is therefore possible to conclude that in 
mechanically-driven turbulence the large superfluid
mesoscale helicity $\mathcal{H}$ is an effect of the normal fluid
via the mutual friction interaction.

\begin{figure}
\centering
\includegraphics[width=0.75\linewidth]{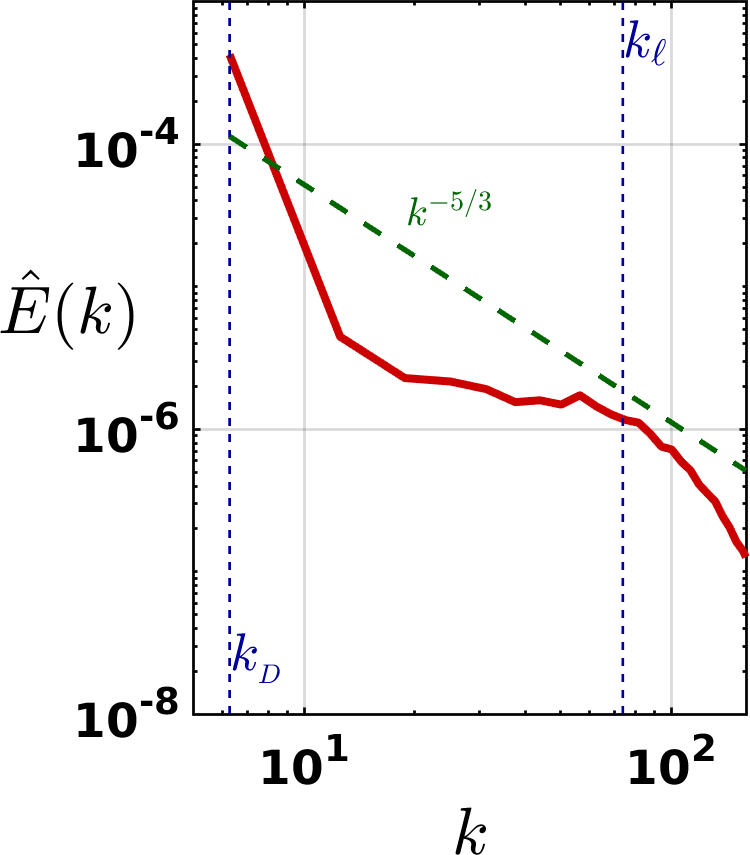}
\caption{
Mechanically-driven turbulence. Superfluid turbulence driven by ABC normal flow:
energy spectrum ${\hat{E}}$ vs wavenumber $k$  at $t=2700 \; \rm{s}$; the dashed green line
shows the slope of the classical Kolmogorov $k^{-5/3}$ scaling, while blue vertical lines
indicate wavenumbers corresponding to the size of the box $D$ and the average intervortex spacing $\ell$.}
\label{fig:ABC_spectra}
\end{figure}

This large value of $\mathcal{H}$ is consistent 
with the computed average magnitude of non-local velocity 
contributions $\boldv^{non}$, in the spirit
of Eq.~\refeq{eq:H3}.
Indeed, if we calculate $\langle |\boldv^{non}| \rangle$ 
in the statistically steady-state 
we obtain $2.1 \times 10^{-2} \rm cm/s$, larger than the corresponding values
evaluated for thermally-driven and injection-driven superfluid turbulence, respectively
$0.45 \times 10^{-2} \rm cm/s$ and $0.96 \times 10^{-2} \rm cm/s$. 
We stress that these computed values of $\langle |\boldv^{non}| \rangle$ for the three 
numerical simulations of superfluid turbulence only show the 
the qualitative behaviour of $\mathcal{H}/(\kappa L)$ across the three distinct systems,
as in Eq.~\refeq{eq:H3} only the {\it tangential component} of $\boldv^{non}$ is averaged
over the vortex tangle. 
To conclude this section, in Fig.~\ref{fig:ABC_spectra} we report the
superfluid energy spectrum ${\hat{E}(k)}$ for this mechanically-driven superfluid
turbulent flow. As a result of the stationary forcing imposed at the largest scale of the flow 
(the box-size $D=1 \; \rm{cm}$), ${\hat{E}(k)}$ is peaked at the highest wavenumber $k_D$, 
producing a much steeper spectrum.

\end{document}